\documentclass{PoS}
\usepackage{lineno}

\title{The Galactic cosmic-ray Sun shadow observed by HAWC}

\ShortTitle{Sun Shadow}

\author{Olivia Enr\'iquez$^a$ and \speaker{Alejandro Lara$^a$} for the
  HAWC collaboration$^b$ \\ 
       \llap{$^a$} Instituto de Geof\'isica, Universidad Nacional Aut\'onoma de M\'exico\\
 \llap{$^b$}For a complete author list, see
\href{http://www.hawc-observatory.org/collaboration/icrc2015.php}{www.hawc-observatory.org/collaboration/icrc2015.php}. \\   
    E-mail: \email{olienriquezrivera@gmail.com}  \email{alara.unam@gmail.com}}

\abstract{
The magnetic field of the Solar corona is difficult to measure
directly. However, indirect observations of the solar corona are
possible using the deficit in flux of cosmic rays coming from the
direction of the Sun. Low-energy cosmic rays  ($\sim$GeV) are deflected by 
the inner magnetic field of the Sun and
the interplanetary magnetic field frozen into the solar wind. 
In contrast, high-energy cosmic rays ($\sim$TeV and above) are
absorbed in the Sun's photosphere producing a
shadow in the Sun's nominal position viewed from Earth. Several
ground-based instruments have observed the effects
of the heliospheric magnetic field on the size of the sun shadow and
its position. The High-Altitude Water Cherenkov Observatory (HAWC) is
an air shower array located in the central region of Mexico that
observes TeV cosmic rays at a rate of about 15 kHz. In this work, we
present preliminary images of the sun shadow from data collected by
HAWC during 2013 and 2014 for different energy ranges. 

}

\FullConference{The 34th International Cosmic Ray Conference,\\
		30 July- 6 August, 2015\\
		The Hague, The Netherlands}

\begin{document}

\section{Introduction}

For an observer on Earth, the Sun and the Moon
block a portion of the Galactic cosmic ray (GCR) flux casting a shadow
equal to their physical angular size~\cite{PhysRev.108.450}, both roughly 0.5$^\circ$ in diameter.
The Moon shadow has been used to calibrate the 
pointing accuracy and angular resolution of air shower detectors as 
early as 1991~\cite{1991PhRvD..43.1735A,1991ICRC....4..460K}. 
An absence of a magnetic field around the Moon allows cosmic rays to travel from the Moon 
to the Earth without large deflections. The Sun has a variable magnetic field which 
can strongly scatter nearby charged cosmic rays 
affecting the geometrical form of the observed Sun shadow. The time
dependence of this field, the so-called ``11-year Solar cycle," 
changes the shape and intensity of the observed Sun shadow.
This allows for the study of the temporal behavior of the Solar magnetic field. 

The Sun shadow was observed by the Tibet 
array~\cite{1993ICRC....4..215A,2000ApJ...541.1051A,2006AdSpR..38..936A,2013PhRvL.111a1101A}
and the ARGO Experiment \cite{2011ICRC...11..158Z}. The authors performed comprehensive studies
with data throughout the 11-year Solar cycle and, using simulations, they showed that it is possible to
obtain information about the Solar magnetic field with the 
GCR Sun shadow \cite{2013PhRvL.111a1101A}.
The Milagro Experiment also reported the observation of the Sun shadow 
\cite{2003ICRC....7.4065X,2011PhDT........70C}. 

The newly finished High-Altitude Water Cherenkov (HAWC) Observatory is able to report on the angular 
observations of 100 GeV to 100 TeV cosmic rays. At these energies, cosmic rays are confined within
the Galaxy and are believe to be of Galactic origin.
Using the first 30 detectors of HAWC, observation of the Moon shadow was in good agreement with 
simulations~\cite{Fiorino:2013}.   
In this work, we present the observations of the Sun shadow with a larger number of detectors 
(between 95 and 111), which allow us to make preliminary estimations of GCR energy.

In Section~\ref{sec:hawc}, we describe HAWC as a GCR detector.
Section~\ref{sec:method} details the method for producing the Sun shadow maps from data. 
HAWC Sun shadow maps, using data acquired during construction, are presented in
Section~\ref{sec:shadow}.
Finally, we conclude and describe future plans for measuring the coronal magnetic field 
using the analysis of the Sun shadow in Section~\ref{sec:summary}.

\section{HAWC} \label{sec:hawc}

The HAWC Observatory is located  
at 4100 m above the sea level 
on the Sierra Negra Volcano in the  central part of Mexico 
(N $18^{\circ}$ 59' 48'', W $97^{\circ}$ 18' 34'').
HAWC consists of 300 water Cherenkov detectors (WCD) covering an area of 22 000 $ m^2$. 
Each WCD is a 
cylindrical container 4.5 m high and 7.3 m in diameter filled with 
$\sim200 000$ liters of filtered water. They are instrumented with 
one  central high quantum efficiency 10'' photo-multipliers (PMT) surrounded by three 
lateral 8''  PMTs. The PMTs face upwards to detect Cherenkov light from relativistic particles, 
produced as secondaries in extensive air showers.
A detailed description of HAWC is  presented in Reference~\cite{Abeysekara2012641}.

The HAWC detector measures the primary particle direction through
accurate temporal and spatial determination of the atmospheric shower front
of secondary particles. The vast 
majority of primary cosmic particles are cosmic rays ($> 99$\%), therefore HAWC is a sensitive 
GCR detector in an energy range going from 100 GeV to 100 TeV.
HAWC can be used to instantaneously monitor more than 2 steradians (sr) of the 
overhead sky with a duty cycle higher than 95\%. This property makes it possible to 
continuously observe the Sun.

The modular construction of HAWC allowed for data acquisition at different 
stages of construction of the array. In this work, we present the data obtained with HAWC-95 and 
HAWC-111, \emph{i. e.} with 95 and 111 WCDs.
We use the number of PMTs hit (nHit) by an air shower as a proxy of the energy of the
primary particle. We present results of three energy-proxy bins:  nHit > 30, 72 and 109. From simulation we determine the primary median energies to be  $\sim$ 2.0, 8.0  and 50 TeV, respectively. The following discussion corresponds to these energy-proxy bins.

The Sun blocks the GCR flux also, but as it has a strong and variable magnetic field,
the produced shadow is less defined. It also changes with time depending on the
Solar activity. We are exploring the possibility of studying the  
Solar magnetic field through the Solar deficit of GCR observed by HAWC.

\section{Method} \label{sec:method}

After the first stage of data reduction, 
the reconstructed arrival direction of the air showers are stored in data maps
that use a HEALpix binning~\cite{2005ApJ...622..759G} of approximately
$0.1^\circ \times  0.1^\circ$ bins Sun-centered coordinate system based on equatorial coordinates. 
Each reconstructed cosmic-ray air shower direction is 
given an azimuthal coordinate of its incident right ascension minus the current Sun right ascension 
and a polar coordinate of its incident declination minus the current Sun declination.
These are referred as $\Delta\alpha$ and $\Delta\delta$.

To obtain the maps of deficits and excesses of air shower events in the sky, 
the data map is compared with a reference map or background map. 
The background map is an estimate of the response of HAWC to an isotropic cosmic 
ray flux. It is important to note that the background map is not itself isotropic due to atmospheric 
effects and the asymmetric shape of the detector~\cite{2015PhDT.........7F}. 
Background maps are produced from the data themselves using the 
method of Direct Integration~\cite{2003ApJ...595..803A} . 

After the background map is created, we apply a smoothing function, a top hat function of 
$2^{\circ}$ radius, in order to draw out correlations in neighboring  bins. 
Intensity maps are calculated by subtracting the background map to the data map. Each bin from these 
maps can be divided by the background counts to obtain relative intensity maps. Finally, we use the 
Li and Ma method~\cite{1983ApJ...272..317L} to find the significance of the relative intensity maps.  

\section{Sun Shadow Maps} \label{sec:shadow}

\begin{figure}
\begin{minipage}{.5\textwidth}
   \includegraphics[width=.99\textwidth]{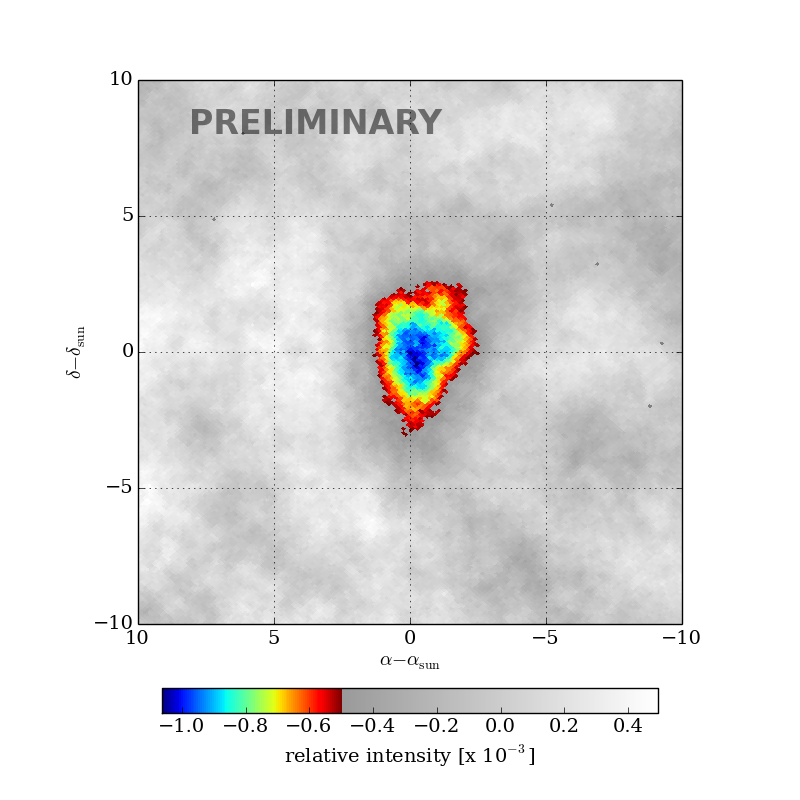} \label{fig:reli30}
\end{minipage}%
\begin{minipage}{.5\textwidth}  
       \includegraphics[width=.99\linewidth]{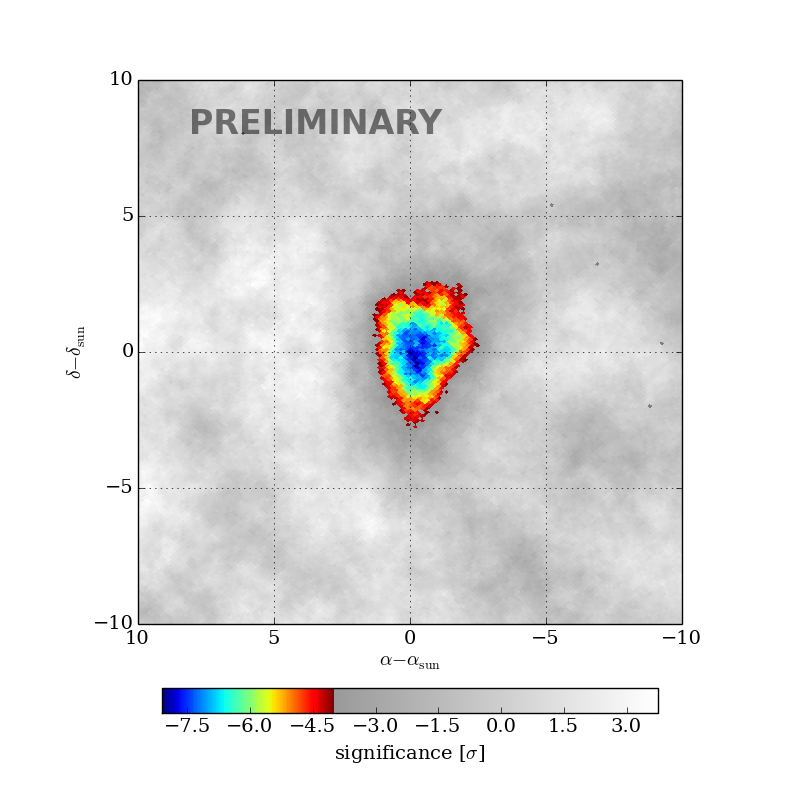}
\end{minipage}
\caption{Relative intensity (left) and significance (right) maps of the Sun shadow with energy-proxy bins > 30 nHit (see text for details)}\label{fig:relisig30}
\end{figure}

\begin{figure}
\begin{minipage}{.5\textwidth}
  \includegraphics[width=.99\textwidth]{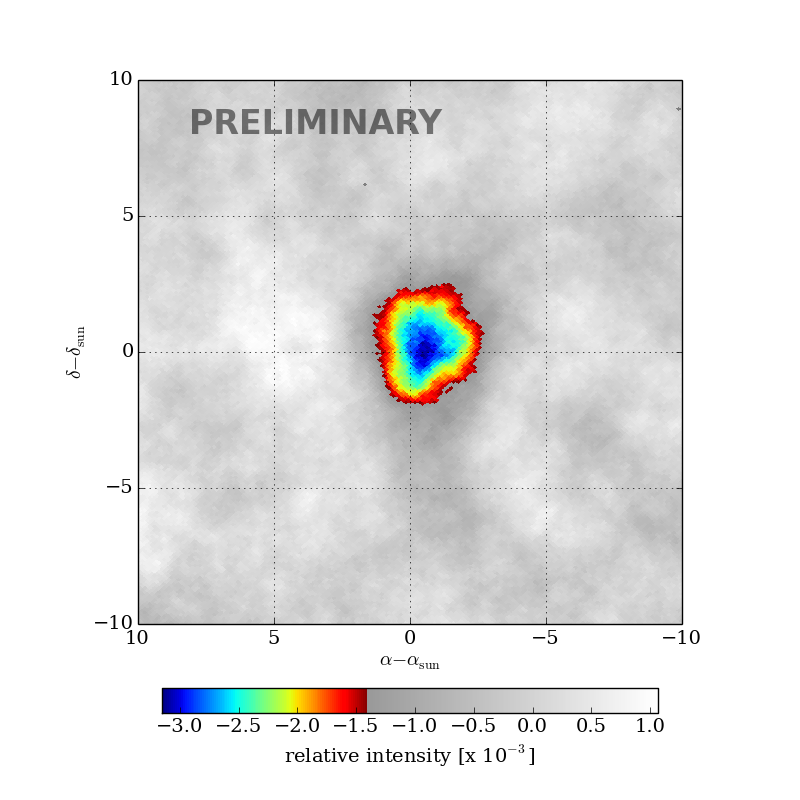} \label{fig:reli72}
  
\end{minipage}%
\begin{minipage}{.5\textwidth}  
       \includegraphics[width=.99\linewidth]{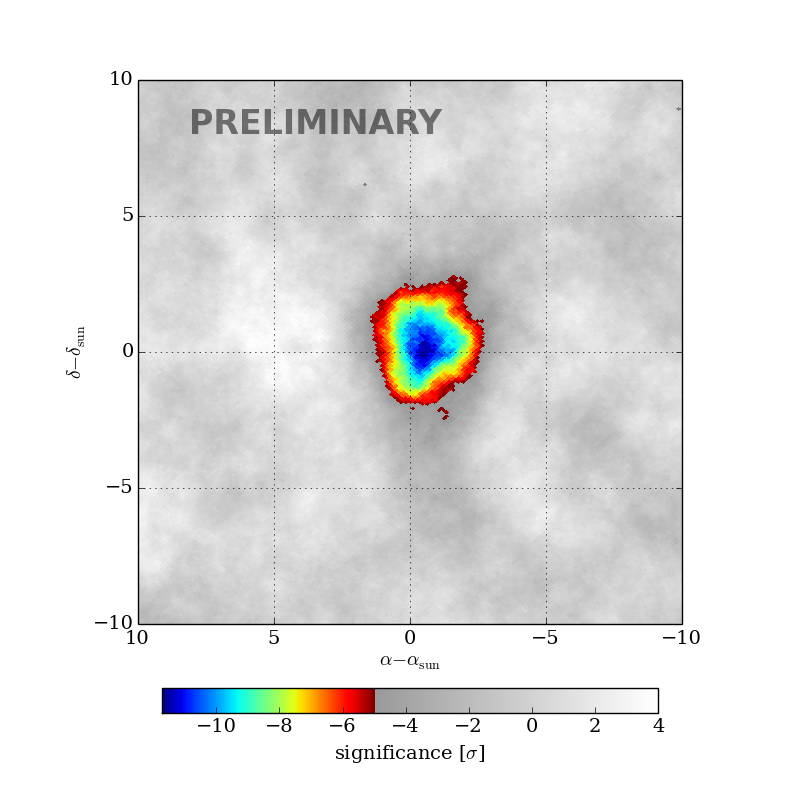}
  \label{fig:test2}
\end{minipage}
\caption{Relative intensity (left) and significance (right) maps of the Sun shadow with energy-proxy bins >72 nHit(see text for details)} \label{fig:relisig72}
\end{figure}

\begin{figure}
\begin{minipage}{.5\textwidth}
 \includegraphics[width=.99\textwidth]{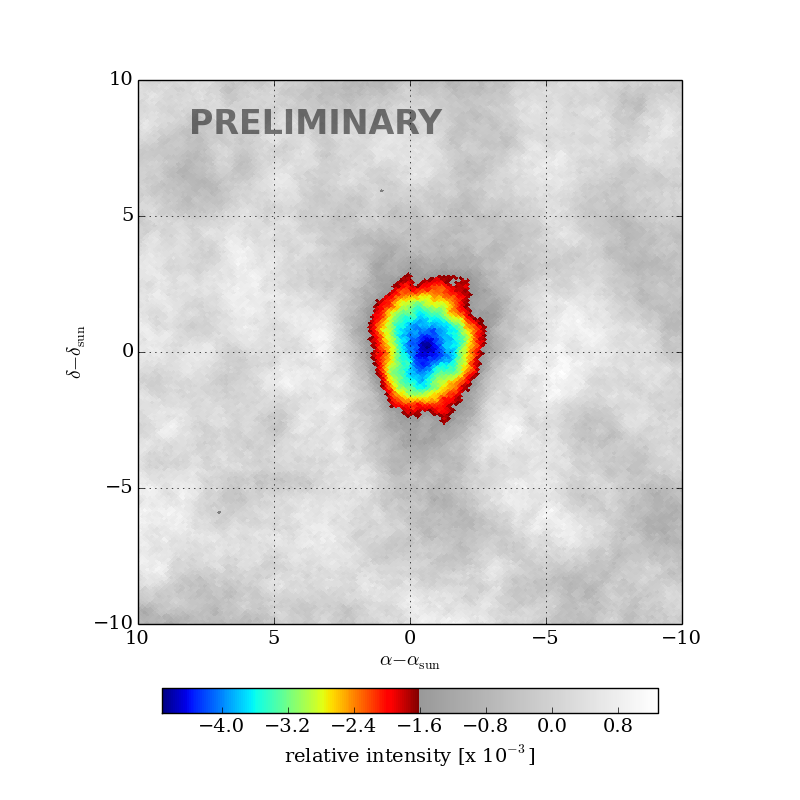}
\end{minipage}%
\begin{minipage}{.5\textwidth}
     \includegraphics[width=.99\linewidth]{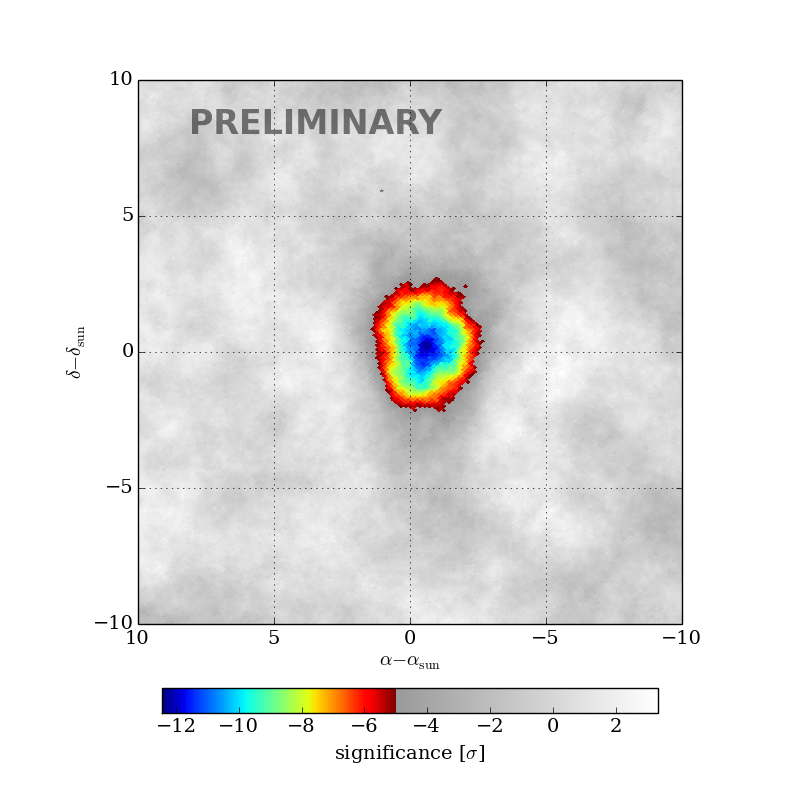}
\end{minipage}
\caption{Relative intensity (left) and significance (right) maps of the Sun shadow with energy cut nHit > 109 (see text for details)}\label{fig:relisig109}

\end{figure}

The significance (relative intensity) maps for the different energy-proxy bins are shown in the right (left) 
panels in Figures~\ref{fig:relisig30}, \ref{fig:relisig72} and \ref{fig:relisig109}. 
We present Sun shadow maps from June 13, 2013 to July 09, 2013, which corresponds to 
HAWC-95/111.  The total duration of the maps is 8121 hrs. 

The size and position of the Sun shadow were obtained by fitting a 2-D asymmetric
Gaussian to the relative intensity maps (Figures~\ref{fig:relisig30}, \ref{fig:relisig72} and 
\ref{fig:relisig109}). Table~\ref{table:res} summarizes the results of the fits.

From Figures~\ref{fig:relisig30}, \ref{fig:relisig72} and \ref{fig:relisig109}, it is clear that the Solar shadow has two components, defined by the energy of the GCR and the 
strength of the Solar magnetic field. 
At low GCR energies the diffusive effect of the Solar magnetic field is strong, causing a 
less defined shadow (Figure~\ref{fig:relisig30}). Whereas at high energies, the GCR flux is blocked by 
the Solar material (Figure~\ref{fig:relisig109}), and the Sun shadow size should approximate to that of 
the Moon. This is also observed in Table~\ref{table:res} where the widths in $\Delta\alpha$ and 
$\Delta\delta$ decrease, \emph{i.e.} the shadow becomes better defined, as the energy-proxy 
bin increases.

In order to quantify the depth of the Sun shadow as the minimum energy-proxy bin changes we have 
plotted the differential relative intensity as a function of angular distance from the 
expected Sun shadow position in Figures 4 and 5. In this case a radial Gaussian fit was applied 
to the data. The amplitude of the fits in the plots indicate that as nHit increases the relative intensity 
reaches higher values. This reflects the fact that high energy GCRs contribute to a greater extent to the 
Sun shadow than low energy GCRs.

\begin{figure}
\centering
\includegraphics[width=.75\textwidth]{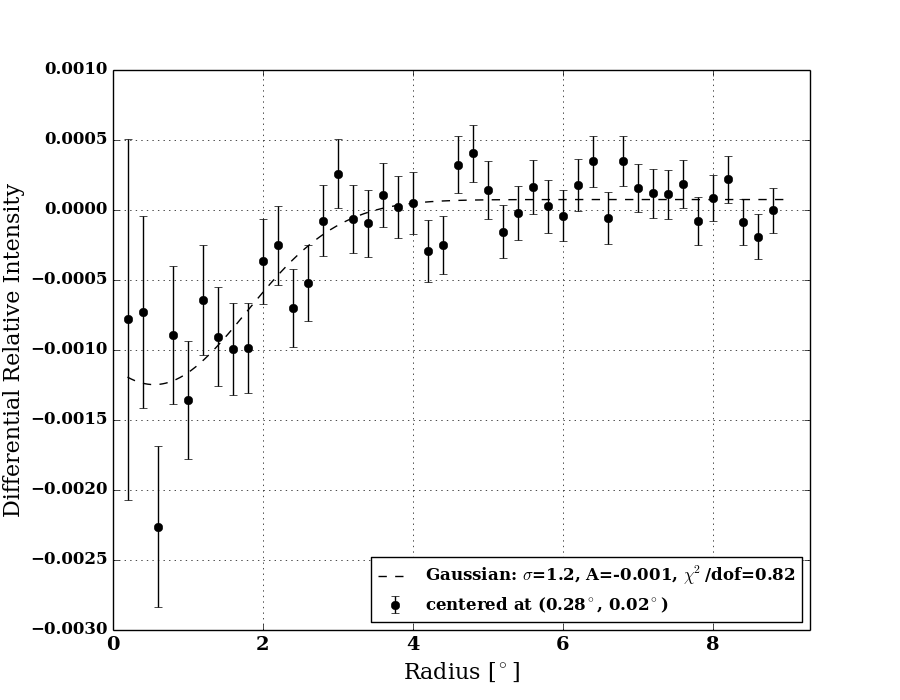}
\caption{The differential relative intensity of GCRs for energy-proxy bin nHit >30,  as a function of radius centered at the Sun position.
A radial Gaussian fit is adjusted to these points (dashed line).
} \label{fig:gauss30}
\end{figure}

\begin{figure}
\begin{minipage}{.5\textwidth}
  \includegraphics[width=.99\textwidth]{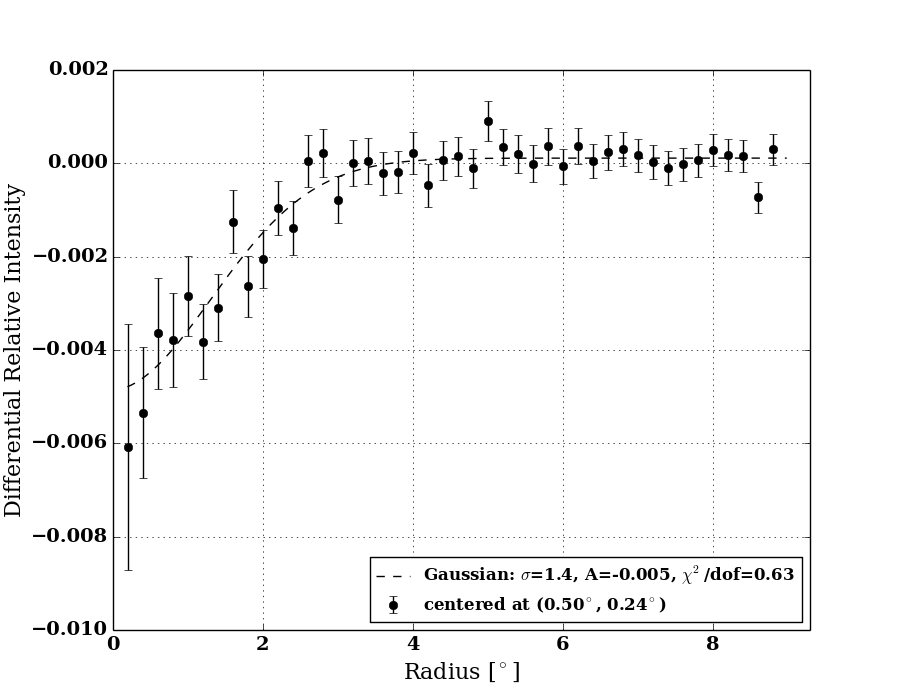}
\end{minipage}%
\begin{minipage}{.5\textwidth}
       \includegraphics[width=.99\linewidth]{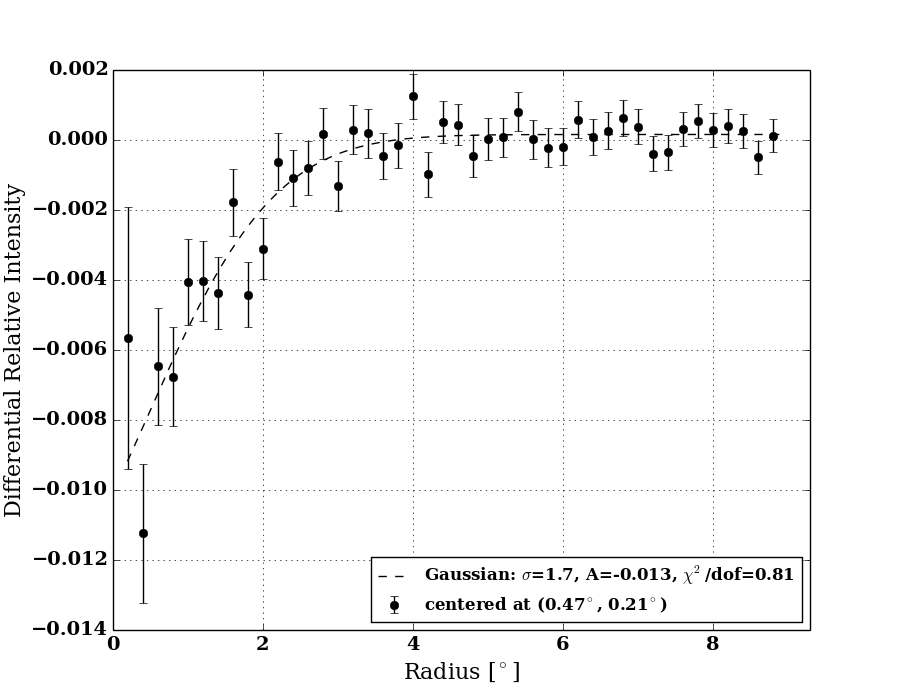}
\end{minipage}
\caption{The differential relative intensity of GCRs for energy-proxy bin nHit >72 (left panel) and nHit >109 (right panel),  as a function of radius and centered at the fit Sun position. A radial Gaussian fit to these points is plotted (dashed line).
}\label{fig:gauss72_109}
\end{figure}

\begin{table} 
\caption{Results of a 2-D asymmetric Gaussian fit of the Sun shadow}  \label{table:res}
\begin{tabular}{crrrrrrrrrrrrrrr}
\hline\hline
 & RA & Dec & & &\\ [1ex]
nHit & $\Delta\alpha$ &  $\Delta\delta$ & Width $\Delta\alpha$  & Width $\Delta\delta$  & Amplitude of fit\\ [1ex]
\hline 

 >30 & 0.28 $\pm$ 0.21 & 0.023 $\pm$ 0.3 & 1.30 $\pm$ 0.22 & 2.18 $\pm$ 0.43& -0.0014 $\pm$ 0.0002\\
 >72 & 0.50 $\pm$ 0.12 & 0.2412 $\pm$ 0.2 & 1.10 $\pm$ 0.13 & 1.59 $\pm$ 0.26&-0.0047 $\pm$ 0.0006  \\
 >109 & 0.47 $\pm$ 0.12 & 0.2184 $\pm$ 0.17 & 1.08 $\pm$ 0.12 & 1.52 $\pm$ 0.25 & -0.0073 $\pm$ 0.0009\\
 \hline
\end{tabular}
\end{table}

\section{Summary} \label{sec:summary} 

The newly finished HAWC observatory detects both, 
electromagnetic and  hadronic air showers created by cosmic radiation. We have used the 
hadronic showers to investigate the effects of the Sun and its magnetic field on the cosmic ray flux. 
Using preliminary data taken during  different stages of the construction 
of the array and which correspond to the maximum phase of the solar cycle 24, we have created maps centered at the Sun position and with a field of
view of 20$^\circ$ in $\Delta\alpha$ and $\Delta\delta$. 
 
We confirmed the existence of a deficit of the GCR flux centered at the Sun. This deficit has been previously observed by other observatories to consist of a strong deficit caused by GCRs directly blocked by the Solar disc (similar in angular size and strength to the Moon shadow) which is weakened by the interaction between GCRs and the Solar magnetic field. This weakening has been correlated with the 11-year Solar cycle.

We plan to use the energy spectrum, time variance (through the solar cycle) and morphology of the Sun shadow maps to  
obtain information about the coronal magnetic field of the Sun . Measurements of 
the strength and large-scale morphology of this field are currently unreachable by direct 
observations at Earth.

\section*{Acknowledgments}
\footnotesize{
We acknowledge the support from: the US National Science Foundation (NSF);
the US Department of Energy Office of High-Energy Physics;
the Laboratory Directed Research and Development (LDRD) program of
Los Alamos National Laboratory; Consejo Nacional de Ciencia y Tecnolog\'{\i}a (CONACyT),
Mexico (grants 260378, 55155, 105666, 122331, 132197, 167281, 167733);
Red de F\'{\i}sica de Altas Energ\'{\i}as, Mexico;
DGAPA-UNAM (grants IG100414-3, IN108713,  IN121309, IN115409, IN111315);
VIEP-BUAP (grant 161-EXC-2011);
the University of Wisconsin Alumni Research Foundation;
the Institute of Geophysics, Planetary Physics, and Signatures at Los Alamos National Laboratory;
the Luc Binette Foundation UNAM Postdoctoral Fellowship program.
}

\bibliography{icrc2015-0716}

\end{document}